\documentclass[fleqn,10pt,twocolumn]{wlscirep}
\usepackage[utf8]{inputenc}
\usepackage[T1]{fontenc}

\providecommand{\bibcommenthead}{}
\providecommand{\burl}[1]{\url{#1}}

\title{Weisfeiler--Lehman subtree encoding for Bayesian optimization
of atomic configurations}

\author[1,2,*]{Akira Kusaba}
\author[3]{Tatoshi Yonemori}
\author[4]{Tetsuji Kuboyama}
\author[1]{Yoshihiro Kangawa}
\affil[1]{Research Institute for Applied Mechanics, Kyushu
University, Fukuoka 816-8580, Japan}
\affil[2]{Institute of Materials and Systems for Sustainability,
Nagoya University, Nagoya 464-8601, Japan}
\affil[3]{Interdisciplinary Graduate School of Engineering Sciences,
Kyushu University, Fukuoka 816-8580, Japan}
\affil[4]{Computer Centre, Gakushuin University, Toshima-ku,
Tokyo 171-8588, Japan}
\affil[*]{kusaba@riam.kyushu-u.ac.jp}

\keywords{Bayesian optimization, atomic configuration search, graph
kernel, Weisfeiler--Lehman, machine-learning interatomic potential,
materials informatics}

\begin{abstract}
The efficiency of Bayesian optimization (BO) of atomic configurations
depends strongly on how configurations are
encoded. We introduce the Weisfeiler--Lehman (WL) subtree kernel,
which views configurations as element-labeled graphs and measures
their similarity by how many local structural patterns they share,
into Bayesian-optimization-based configuration search. Because this
kernel is reproduced as the plain
inner product of explicit features---$L^2$-normalized histograms of
local topological patterns---introducing it reduces to introducing
the corresponding features:
the encoding enters existing BO frameworks as an ordinary
descriptor. In a benchmark ground-state
configuration search of cubic BC$_2$N evaluated with a universal
machine-learning interatomic potential, the WL encoding reached the
ground state almost immediately after a shared random initialization
of 100 samples in every one of five independent rounds ($108\pm5$
evaluations on average), whereas the one-hot baseline required
$280\pm122$ evaluations; the WL-driven
sampler first exhausted the degenerate ground-state group
and then discovered the metastable degenerate groups from the bottom
up, in order of increasing energy.
\end{abstract}

\begin{document}

\flushbottom
\maketitle
\thispagestyle{empty}

\section*{Introduction}

In multicomponent crystals and alloys, the arrangement of atomic species
on a fixed lattice---the atomic configuration---can govern stability
and functional properties, and identifying the most stable
configuration is a combinatorial optimization problem: even a modest
supercell of a few tens of sites with a fixed composition spans on
the order of $10^{5}$ or more
symmetry-distinct candidates, far beyond exhaustive first-principles
evaluation. (This search for specific stable patterns is
complementary to the statistical modeling of configurational
disorder.\cite{Zunger1990,Kasamatsu2022,Kasamatsu2023}) One well-established strategy is
to fit a lightweight on-lattice energy model---the cluster
expansion\cite{Sanchez1984} with mature software such as ATAT and
icet\cite{vandeWalle2002,Angqvist2019}---to first-principles
energies, and then to sample the fitted model
massively by Monte Carlo annealing, genetic algorithms, or
exhaustive enumeration, at negligible cost per sample. Bayesian
optimization (BO) with a Gaussian-process (GP) surrogate takes the
opposite route: it retains the expensive evaluator---density
functional theory (DFT) or a machine-learning interatomic potential
(MLIP)---and instead economizes on the number of evaluations,
spending them only on sequentially selected informative
candidates.\cite{Ueno2016,Yamashita2018,Lookman2019,Terayama2021}
PyAPX\cite{Kusaba2025,Kusaba2022,Kawka2024,Hara2025} is an open-source
toolkit that automates this BO workflow: candidate configurations are
encoded into feature vectors and passed to the PHYSBO
engine,\cite{Motoyama2022} which performs GP regression with a random
feature map and selects candidates by Thompson sampling, while
energies are supplied by
DFT or MLIPs, including the recent universal
models,\cite{Deng2023,Wood2025} through a pluggable
evaluator interface.

The sample efficiency of BO hinges on the encoding, i.e., on the
similarity structure the GP assumes between configurations. The
toolkit provides a one-hot encoding of site occupancies and,
from our previous work,\cite{Kusaba2025} neighboring-atom variants
(NA, NAmod) that describe the environment of each site by weighted
counts of the neighboring species---its local composition---and, in
NAmod, additionally by the anisotropy of that environment. NAmod showed clearly superior convergence over one-hot for
a two-dimensional system, a monolayer h-BCN sheet, but lost its
advantage on a more challenging three-dimensional problem, cubic
BC$_2$N:\cite{Solozhenko2001} the local-environment
encodings were merely comparable to one-hot in the early stage of
sampling, and one-hot settled at a slightly lower energy level in the
later stage.\cite{Kusaba2025} A plausible reason is the assumption
built into such numeric summaries---that environments of similar
local composition (and its anisotropy) contribute similar
energies---which is
consistent with the bond-preference-dominated energetics of h-BCN
but appears to be violated in c-BC$_2$N.\cite{Kusaba2025} In this
work we therefore
represent each local environment not by its composition but as an
exact, discrete graph pattern, and describe a configuration simply
by counting which patterns occur. Only locality is then assumed:
the energy is a sum of per-pattern contributions, and no relation
is presumed between distinct patterns.

\section*{Weisfeiler--Lehman subtree encoding}

\begin{figure*}[!tb]
\centering
\includegraphics[width=0.95\textwidth]{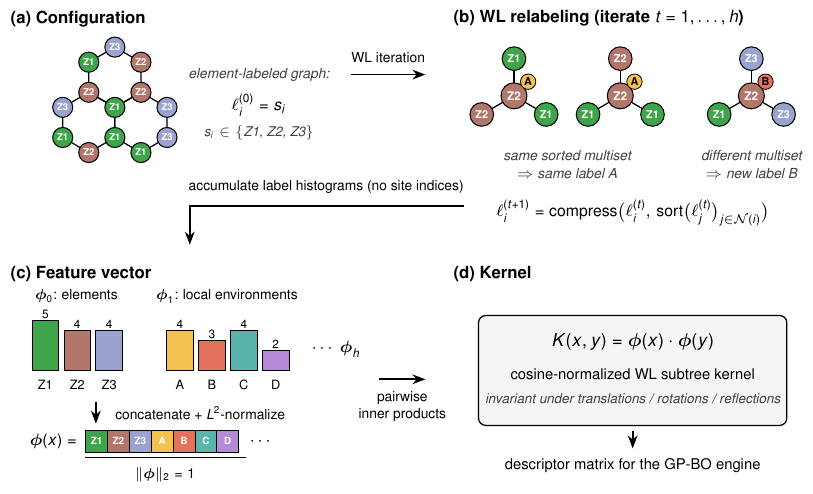}
\caption{WL subtree encoding of an atomic
configuration. (a) Element-labeled graph of a
configuration. (b) One step of the WL relabeling, Eq.~(\ref{eq:wl}):
sites with identical (own label, sorted neighbor multiset) pairs
receive the same new label. (c) Per-level label histograms,
concatenated and $L^2$-normalized into the feature vector,
Eq.~(\ref{eq:feat}). (d) The plain inner product of the features,
Eq.~(\ref{eq:kernel}), realizes the WL kernel; the feature matrix
enters the GP-BO engine as a descriptor.}
\label{fig:wl}
\end{figure*}

Exactly such an encoding is provided by the
Weisfeiler--Lehman (WL) subtree construction,\cite{Weisfeiler1968,
Shervashidze2011} a standard device in graph
learning.\cite{Kriege2020} The same relabeling underlies modern graph
neural networks, whose message-passing expressivity is bounded by the
WL test;\cite{XuGNN2019,Morris2019} the WL encoding can thus be
regarded as a training-free kernel counterpart of crystal-graph
networks.\cite{Xie2018} WL-based kernels have recently been applied
in materials and chemistry contexts, e.g., Gaussian-process regression
of adsorption enthalpies on transition-metal alloys\cite{Xu2022} and
molecular property prediction and optimization,\cite{Griffiths2023}
but, to our knowledge, they have not been used for
BO-based atomic-configuration search. The closest approach in spirit is
configurational BO with cluster-expansion correlation functions as
features;\cite{Seko2020} the WL encoding differs decisively in that
its basis is not designed by a symmetry analysis of the lattice with a
truncated many-body expansion, but emerges from the candidate set
itself, each label encoding a complete neighborhood pattern up to a
prescribed depth---effectively all interaction orders within that
range---while requiring only the bond list as
input. Continuous local-environment descriptors such as
SOAP\cite{Bartok2013} and the atomic cluster
expansion\cite{Drautz2019} serve the same purpose---representing
local atomic environments as machine-learning features---for
off-lattice geometries; the WL construction is their natural
graph-based analogue
for the discrete occupancy problem on a fixed lattice. A
configuration is regarded as a
labeled graph (Fig.~\ref{fig:wl}): sites are vertices, bonds between
neighboring sites are edges, and the initial label of vertex $i$ is
its atomic species, $\ell_i^{(0)} = s_i$. The WL relabeling
\begin{equation}
\ell_i^{(t+1)} \;=\;
\mathrm{compress}\!\left(\ell_i^{(t)},\,
\mathrm{sort}\bigl(\ell_j^{(t)}\bigr)_{j \in \mathcal{N}(i)}\right)
\label{eq:wl}
\end{equation}
where $\mathcal{N}(i)$ is the neighbor list of site $i$, taken
as a multiset so that repeated neighbors are counted with
multiplicity,
assigns a common new label to identical pairs of (own label, sorted
multiset of neighbor labels) and distinct new labels to distinct
pairs; one and the same mapping is shared by all sites of all
candidates. After $t$ iterations, therefore,
$\ell_i^{(t)}$ identifies the topology and chemical decoration of the
depth-$t$ neighborhood (subtree) rooted at site $i$. For each level
$t = 0,\dots,h$ we accumulate the label histogram
$\phi_t(x)$ over all sites of configuration $x$, concatenate the
levels, and normalize,
\begin{equation}
\phi(x) = \frac{[\phi_0(x);\,\dots;\,\phi_h(x)]}
{\lVert[\phi_0(x);\dots;\phi_h(x)]\rVert_2},
\label{eq:feat}
\end{equation}
\begin{equation}
K(x,y) = \phi(x)\!\cdot\!\phi(y),
\label{eq:kernel}
\end{equation}
so that the plain inner product of the features equals the
cosine-normalized WL subtree kernel. This explicitness is the key to a
minimal implementation: $\phi$ is simply handed to the BO engine as
a descriptor matrix, and the entire downstream
pipeline---column-wise feature standardization,
Gaussian kernel with random feature map, and Thompson
sampling---operates unchanged. When the engine's Gaussian
kernel---the covariance function of the GP surrogate---acts
directly on these features, because $\lVert\phi\rVert_2=1$, it is
a monotone transform of the WL similarity,
$\exp\{[K(x,y)-1]/\sigma^2\}$, and the similarity structure of the
WL kernel is carried into the surrogate model
intact. In the experiments below, however, the engine's
default standardization precedes the kernel, so the realized
covariance is a feature-weighted variant of this form; the symmetry
invariance of the representation (equivalent configurations receive
identical features) is unaffected. A control experiment with the
standardization disabled, so that the kernel acts directly on $\phi$,
reproduced the WL result reported below ($105\pm6$ versus $108\pm5$
evaluations to the ground state), confirming that this does not
affect the conclusions.

The physical content of this representation differs fundamentally
from that of the existing encodings (one-hot, NA, and NAmod): those
descriptors record \textit{which
species occupies which site}, whereas the WL features record only
\textit{how often each local environment occurs}---a configuration is
reduced to the multiset of its labeled local structures, with all site
indices discarded. This is precisely the information that determines
the energy whenever the energy is a sum of local contributions, and
nothing more. The encoding thus has two properties of practical
value for configuration search. (i) \textit{Symmetry invariance}:
$\phi$ depends only on the multiset of local structures, so
configurations related by translations, rotations, or reflections of
the lattice (graph automorphisms) receive identical features;
equivalent candidates collapse to a single point in feature space,
shrinking the effective search space. (ii) \textit{Site-count
independence}: the dimension is set by the vocabulary of distinct
local patterns rather than by the number of sites. In the implementation, the shared mapping of
Eq.~(\ref{eq:wl}) is realized by performing the label compression
jointly over all candidates (a single \texttt{unique} over the
pooled keys). In PyAPX the new encoding is exposed in the
established input format:
the user sets \texttt{ENCODE\_TYPE = WL} and, optionally,
\texttt{WL\_H} (the iteration depth $h$) in \texttt{apx.in}, and
supplies the same \texttt{NEIGHBOR\_SITES} card
already used by NA/NAmod.

\section*{Results}

We applied the WL encoding to the stable-configuration search
of cubic BC$_2$N in a $2\times2\times2$ supercell of the primitive
zinc-blende/diamond cell: 16
sites with four bonded neighbors each,
occupied by 4B$+$8C$+$4N. Following our previous
work,\cite{Kusaba2025} site 1 was fixed to C to reduce
translationally equivalent configurations, giving a candidate pool
of $\binom{15}{4}\binom{11}{4} = 450{,}450$
configurations. Total energies
were evaluated by the universal MLIP UMA (model
\texttt{uma-m-1p1})\cite{Wood2025} through the pluggable evaluator
interface. In the following, the \textit{ground state} refers to
the lowest-energy configuration found in this study, to which all
ten runs described below converged. The BO protocol was 100 random samples followed by 900
Bayesian samples (PHYSBO, Thompson sampling, 3000 random bases), and
we compared the WL encoding ($h=1$) with the one-hot baseline over five
independent rounds. Within each round the 100 random samples were
identical between the two encodings, so
any within-round performance difference is attributable to the
encoding rather than to the random draw of the initial samples; the
initializations of different rounds share no samples. The
depth $h=1$ was chosen deliberately: for a 16-site cell the $h=1$
vocabulary comprises only $\sim$45 local environments, whereas $h=2$
admits up to $\sim$$1.2\times10^{5}$ combinatorially possible labels,
and such over-discrimination sparsifies the similarity structure.

\begin{figure}[!tb]
\centering
\includegraphics[width=\linewidth]{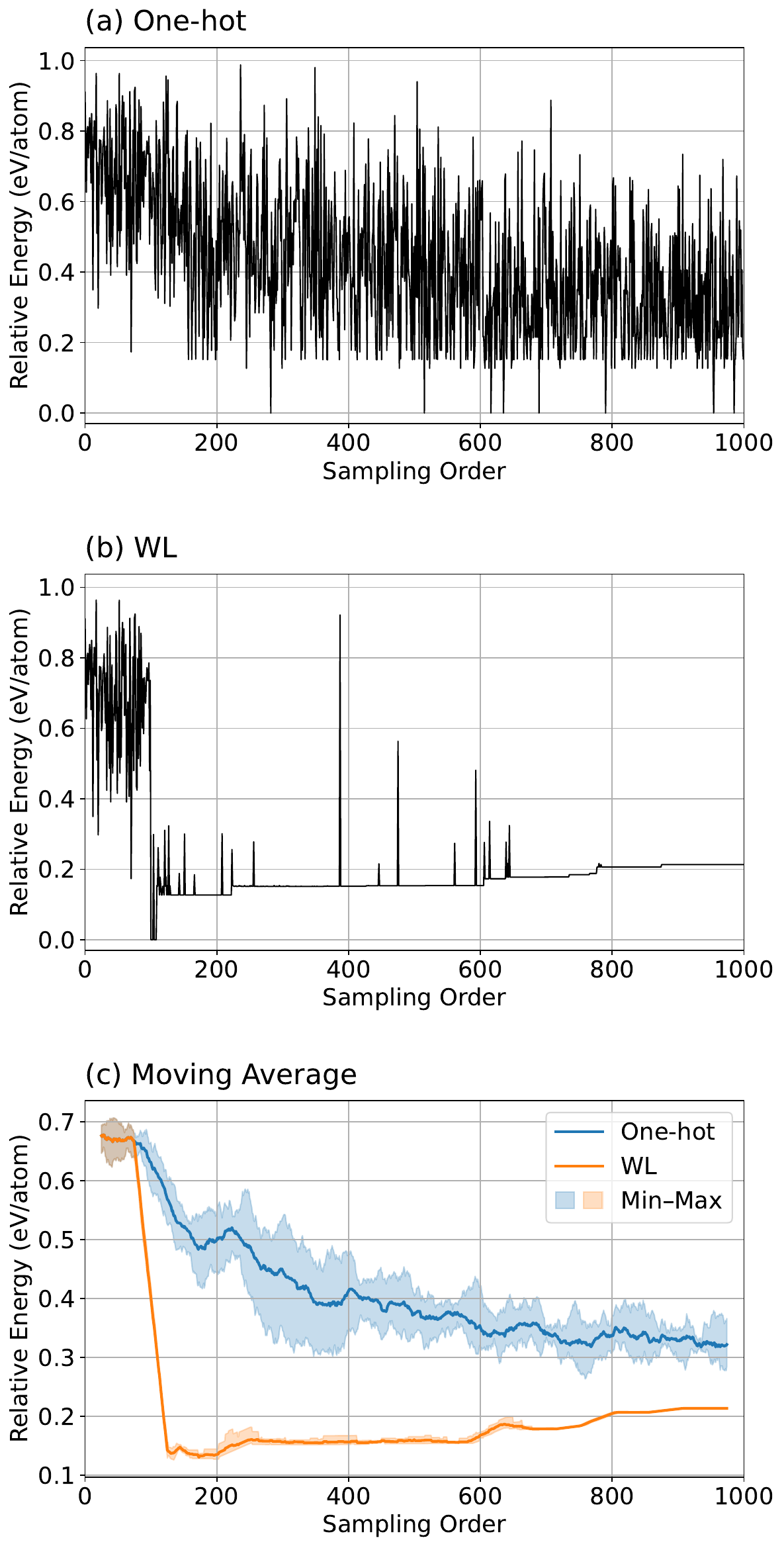}
\caption{Configuration search of cubic BC$_2$N ($2\times2\times2$
zinc-blende/diamond supercell, 4B$+$8C$+$4N) with UMA MLIP energies,
for the
one-hot and WL ($h=1$) encodings with a shared 100-sample random
initialization. (a),(b) Raw sampling histories of a representative
round; energies are per atom relative to the common ground state.
(c) Moving average (window 51) of each round's history, summarized
over the five independent rounds: solid lines show the mean across
rounds, and shaded bands the round-to-round minimum--maximum
range.}
\label{fig:bc2n}
\end{figure}

All ten runs (two encodings $\times$ five rounds) reached the same
ground state, of which the candidate pool contains eight
symmetry-equivalent copies. The WL encoding
reached the ground state after $108\pm5$ evaluations (mean $\pm$ standard
deviation over the five rounds; per-round
values 101--115)---in every round within at most 15 Bayesian samples
after the random stage---whereas one-hot required $280\pm122$ evaluations
(range 150--479), 2.6 times more on average and with a far larger
round-to-round spread. In the representative round shown in
Fig.~\ref{fig:bc2n}, WL hit the ground state at evaluation 101, i.e., with
its very first Bayesian sample, versus 283 for one-hot. The advantage is
not limited to the first hit: in that round the mean total energy
over the 900 Bayesian samples was $-136.94$~eV for WL versus
$-133.26$~eV for one-hot, a gap of 0.23~eV/atom, and across the five
rounds the moving averages remain separated---the min--max band of WL
lies entirely below that of one-hot throughout the Bayesian stage
[Fig.~\ref{fig:bc2n}(c)].

The degenerate structure of the low-energy spectrum makes visible
the mechanism behind the WL advantage. In
the round of Fig.~\ref{fig:bc2n}, both
encodings independently discovered exactly the same set of eight
candidate configurations for the ground state, mutually
confirming its
eightfold degeneracy; the discovery patterns, however, are
contrasting. The WL run consumed the eight copies almost consecutively
(evaluations 101--109, eight of nine samples): since all copies
coincide at a single point in the symmetry-invariant feature space,
the acquisition function drains the whole group at once. The one-hot run instead
rediscovered the copies sporadically over evaluations 283--986,
treating each as an unrelated candidate and spending a large part of
its budget re-finding the same physical structure. After exhausting
the ground-state group, the WL sampler sweeps the degenerate excited
groups upward in a stair-like fashion, seen in
Fig.~\ref{fig:bc2n}(b) as flat stretches of consecutive samples at
$\Delta E \approx 0.13$, 0.15, 0.17, and
0.20~eV/atom---Thompson sampling systematically enumerating
the low-energy configurations group by group---whereas the one-hot
moving average only reaches
$\sim$0.3~eV/atom within the 1000-sample budget in every round
[Fig.~\ref{fig:bc2n}(c)].

The bonding in the sampled configurations makes the difference
sharper still. B--B and N--N bonds are known to be strongly
unfavorable in BCN systems, and the ground state found here contains
none. Although configurations free of B--B and N--N bonds make up
only 2.2\% of the pool (9{,}726 of 450{,}450), in the WL runs 98\% of
the first 100 Bayesian samples (evaluations 101--200) were such
configurations, and the fraction stayed at essentially 100\%
thereafter (five-round average); in the one-hot runs the same fraction
was 19\% over evaluations 101--200 and reached only 55\% in the final
stretch (evaluations 701--1000). The count of each bond type is a
product of two site-occupancy variables in the one-hot features,
whereas in the $h=1$ WL features it is built in from the outset as a
linear function of the label frequencies. Bond counting alone does not
fix the ordering, however: the five excited groups just above the
ground state ($\Delta E \approx 0.13$--0.15~eV/atom) have identical
bond counts (B--B $=$ N--N $=0$, B--N $=10$, C--C $=10$,
C--B $=$ C--N $=6$), and what distinguishes them is the combination
of local motifs.

\section*{Discussion}

The effectiveness of the WL encoding admits a simple physical
reading. Whenever the energy is a sum of contributions each
determined by a site's depth-$h$ environment---of any interaction
order within that environment, with pairwise bond counting as the
simplest case---the energy is exactly a linear functional of the
level-$h$ histogram, and the WL feature vector is a sufficient
statistic for it; when the energy departs from such a sum only by a
small nonlocal remainder, the feature vector remains close to
sufficient, so
the GP posterior aligns with the physics after few observations. Symmetry invariance additionally
folds all equivalent configurations onto single feature points, and
the absence of site-index information prevents the surrogate from
overfitting to particular lattice positions. Because the WL dimension
tracks the diversity of local environments rather than the site count,
the advantage is expected to grow with system size.

Three limitations deserve note. First, although the WL
vocabulary---and hence the feature dimension---is independent of the
site count, it grows with the number of chemical species, the
coordination number, and the depth; at $h=1$
it stays small (tens of environments in the present system), but for
large pools with $h\ge2$ the feature matrix (candidates $\times$
vocabulary), if stored densely with explicit zeros, can become
memory-limited. We recommend $h=1$ in any case, given the
over-discrimination at larger depths noted above; until sparse-feature
support is available, restricting the search to a random subset of
the pool or compressing the features by principal component analysis
may also serve. Second, the WL test cannot
distinguish certain nonisomorphic graphs in principle; no adverse
effect of such degeneracy was observed in the present search. Third, when relaxation
introduces long-range or nonadditive energy components, the
local-structure histogram is no longer a sufficient statistic; the
c-BC$_2$N$+$UMA result indicates that the encoding remains effective
in practice, but level-dependent weighting of the WL histograms is an
interesting refinement.

The repeated evaluation of the eightfold-degenerate ground state also
points to a natural extension: generating the candidate pool with
SHRY,\cite{Prayogo2022} which enumerates only symmetry-inequivalent
configurations by canonical augmentation, would remove such redundancy
at the source for any encoding. The combination of a
SHRY-reduced pool with the WL encoding is complementary rather than
redundant: the pool removes global symmetry copies, while the
WL features keep physically similar (but inequivalent) orderings close
in feature space, and the WL dimension---set by the vocabulary of
local patterns rather than the site count---keeps the regression
problem essentially unchanged as the supercell grows. We therefore
expect the combination of SHRY and the WL encoding to be effective
for larger cells; its demonstration is left for future work.

\section*{Conclusions}

In summary, we introduced the WL subtree kernel into BO-based
atomic-configuration search. Because this kernel is the plain inner product
of explicit, finite-dimensional features---$L^2$-normalized
histograms of local patterns---it could be implemented as an
ordinary descriptor: a drop-in alternative to the existing
encodings, requiring no change to the machinery that performs the
surrogate regression and candidate
selection. In the ground-state search of cubic
BC$_2$N with a universal MLIP, it reached the ground state almost
immediately after a shared 100-sample random initialization in every
one of five
independent rounds ($108\pm5$ total evaluations versus $280\pm122$
for one-hot), sampled configurations averaging
0.23~eV/atom lower throughout the run, and systematically enumerated
the eightfold-degenerate ground-state group together with the
low-lying degenerate excited groups. Because the encoding is
symmetry-invariant and its dimension does not scale with the number
of sites, it combines naturally with symmetry-reduced candidate
pools, extending configuration search to supercells that were
impractical with the previous workflow. The implementation is
available in the PyAPX repository.

\section*{Data availability}
The data that support the findings of this study are available from
the corresponding author upon reasonable request. The PyAPX toolkit,
including the WL encoding, is publicly available on GitHub at
\url{https://github.com/a-ksb/PyAPX} (v1.1.1).

\section*{Acknowledgements}
This work was partially supported by JSPS KAKENHI (grant numbers
JP23K28151, JP24K17619, JP24H00432); JST BOOST (grant number
JPMJBY24C3); and Collaborative Research Program of Research Institute
for Applied Mechanics, Kyushu University. The computation was
performed using Research Center for Computational Science, Okazaki,
Japan (Project: 26-IMS-C120).


\begin{thebibliography}{10}
\expandafter\ifx\csname url\endcsname\relax
  \def\url#1{\burl{#1}}\fi
\expandafter\ifx\csname urlprefix\endcsname\relax\def\urlprefix{URL }\fi
\providecommand{\bibinfo}[2]{#2}
\providecommand{\eprint}[2][]{\url{#2}}
\providecommand{\doi}[1]{\url{https://doi.org/#1}}
\bibcommenthead

\bibitem{Zunger1990}
\bibinfo{author}{Zunger, A.}, \bibinfo{author}{Wei, S.-H.},
  \bibinfo{author}{Ferreira, L.~G.} \& \bibinfo{author}{Bernard, J.~E.}
\newblock \bibinfo{title}{Special quasirandom structures}.
\newblock \emph{\bibinfo{journal}{Phys. Rev. Lett.}}
  \textbf{\bibinfo{volume}{65}}, \bibinfo{pages}{353--356}
  (\bibinfo{year}{1990}).

\bibitem{Kasamatsu2022}
\bibinfo{author}{Kasamatsu, S.} \emph{et~al.}
\newblock \bibinfo{title}{Facilitating ab initio configurational sampling of
  multicomponent solids using an on-lattice neural network model and active
  learning}.
\newblock \emph{\bibinfo{journal}{J. Chem. Phys.}}
  \textbf{\bibinfo{volume}{157}}, \bibinfo{pages}{104114}
  (\bibinfo{year}{2022}).

\bibitem{Kasamatsu2023}
\bibinfo{author}{Kasamatsu, S.}, \bibinfo{author}{Motoyama, Y.},
  \bibinfo{author}{Yoshimi, K.} \& \bibinfo{author}{Aoyama, T.}
\newblock \bibinfo{title}{Configuration sampling in multi-component
  multi-sublattice systems enabled by ab initio configuration sampling toolkit
  ({abICS})}.
\newblock \emph{\bibinfo{journal}{Sci. Technol. Adv. Mater.: Methods}}
  \textbf{\bibinfo{volume}{3}}, \bibinfo{pages}{2284128}
  (\bibinfo{year}{2023}).

\bibitem{Sanchez1984}
\bibinfo{author}{Sanchez, J.~M.}, \bibinfo{author}{Ducastelle, F.} \&
  \bibinfo{author}{Gratias, D.}
\newblock \bibinfo{title}{Generalized cluster description of multicomponent
  systems}.
\newblock \emph{\bibinfo{journal}{Physica A}} \textbf{\bibinfo{volume}{128}},
  \bibinfo{pages}{334--350} (\bibinfo{year}{1984}).

\bibitem{vandeWalle2002}
\bibinfo{author}{van~de Walle, A.}, \bibinfo{author}{Asta, M.} \&
  \bibinfo{author}{Ceder, G.}
\newblock \bibinfo{title}{The alloy theoretic automated toolkit: A user guide}.
\newblock \emph{\bibinfo{journal}{Calphad}} \textbf{\bibinfo{volume}{26}},
  \bibinfo{pages}{539--553} (\bibinfo{year}{2002}).

\bibitem{Angqvist2019}
\bibinfo{author}{{\r{A}}ngqvist, M.} \emph{et~al.}
\newblock \bibinfo{title}{{ICET} -- a {P}ython library for constructing and
  sampling alloy cluster expansions}.
\newblock \emph{\bibinfo{journal}{Adv. Theory Simul.}}
  \textbf{\bibinfo{volume}{2}}, \bibinfo{pages}{1900015}
  (\bibinfo{year}{2019}).

\bibitem{Ueno2016}
\bibinfo{author}{Ueno, T.}, \bibinfo{author}{Rhone, T.~D.},
  \bibinfo{author}{Hou, Z.}, \bibinfo{author}{Mizoguchi, T.} \&
  \bibinfo{author}{Tsuda, K.}
\newblock \bibinfo{title}{{COMBO}: An efficient {B}ayesian optimization library
  for materials science}.
\newblock \emph{\bibinfo{journal}{Mater. Discov.}}
  \textbf{\bibinfo{volume}{4}}, \bibinfo{pages}{18--21} (\bibinfo{year}{2016}).

\bibitem{Yamashita2018}
\bibinfo{author}{Yamashita, T.} \emph{et~al.}
\newblock \bibinfo{title}{Crystal structure prediction accelerated by
  {B}ayesian optimization}.
\newblock \emph{\bibinfo{journal}{Phys. Rev. Mater.}}
  \textbf{\bibinfo{volume}{2}}, \bibinfo{pages}{013803} (\bibinfo{year}{2018}).

\bibitem{Lookman2019}
\bibinfo{author}{Lookman, T.}, \bibinfo{author}{Balachandran, P.~V.},
  \bibinfo{author}{Xue, D.} \& \bibinfo{author}{Yuan, R.}
\newblock \bibinfo{title}{Active learning in materials science with emphasis on
  adaptive sampling using uncertainties for targeted design}.
\newblock \emph{\bibinfo{journal}{npj Comput. Mater.}}
  \textbf{\bibinfo{volume}{5}}, \bibinfo{pages}{21} (\bibinfo{year}{2019}).

\bibitem{Terayama2021}
\bibinfo{author}{Terayama, K.}, \bibinfo{author}{Sumita, M.},
  \bibinfo{author}{Tamura, R.} \& \bibinfo{author}{Tsuda, K.}
\newblock \bibinfo{title}{Black-box optimization for automated discovery}.
\newblock \emph{\bibinfo{journal}{Acc. Chem. Res.}}
  \textbf{\bibinfo{volume}{54}}, \bibinfo{pages}{1334--1346}
  (\bibinfo{year}{2021}).

\bibitem{Kusaba2025}
\bibinfo{author}{Kusaba, A.} \emph{et~al.}
\newblock \bibinfo{title}{{PyAPX}: python toolkit for atomic configuration
  pattern exploration}.
\newblock \emph{\bibinfo{journal}{Sci. Rep.}}  (\bibinfo{year}{2026}).
\newblock \bibinfo{note}{{https://doi.org/10.1038/s41598-026-66072-5}}.

\bibitem{Kusaba2022}
\bibinfo{author}{Kusaba, A.}, \bibinfo{author}{Kangawa, Y.},
  \bibinfo{author}{Kuboyama, T.} \& \bibinfo{author}{Oshiyama, A.}
\newblock \bibinfo{title}{Exploration of a large-scale reconstructed structure
  on {GaN}(0001) surface by {B}ayesian optimization}.
\newblock \emph{\bibinfo{journal}{Appl. Phys. Lett.}}
  \textbf{\bibinfo{volume}{120}}, \bibinfo{pages}{021602}
  (\bibinfo{year}{2022}).

\bibitem{Kawka2024}
\bibinfo{author}{Kawka, K.} \emph{et~al.}
\newblock \bibinfo{title}{Augmentation of the electron counting rule with
  {I}sing model}.
\newblock \emph{\bibinfo{journal}{J. Appl. Phys.}}
  \textbf{\bibinfo{volume}{135}}, \bibinfo{pages}{225302}
  (\bibinfo{year}{2024}).

\bibitem{Hara2025}
\bibinfo{author}{Hara, T.} \emph{et~al.}
\newblock \bibinfo{title}{Exploration of stable atomic configurations in
  graphene-like {BCN} systems by density functional theory and {B}ayesian
  optimization}.
\newblock \emph{\bibinfo{journal}{Cryst. Growth Des.}}
  \textbf{\bibinfo{volume}{25}}, \bibinfo{pages}{6719--6726}
  (\bibinfo{year}{2025}).

\bibitem{Motoyama2022}
\bibinfo{author}{Motoyama, Y.} \emph{et~al.}
\newblock \bibinfo{title}{Bayesian optimization package: {PHYSBO}}.
\newblock \emph{\bibinfo{journal}{Comput. Phys. Commun.}}
  \textbf{\bibinfo{volume}{278}}, \bibinfo{pages}{108405}
  (\bibinfo{year}{2022}).

\bibitem{Deng2023}
\bibinfo{author}{Deng, B.} \emph{et~al.}
\newblock \bibinfo{title}{{CHGNet} as a pretrained universal neural network
  potential for charge-informed atomistic modelling}.
\newblock \emph{\bibinfo{journal}{Nat. Mach. Intell.}}
  \textbf{\bibinfo{volume}{5}}, \bibinfo{pages}{1031--1041}
  (\bibinfo{year}{2023}).

\bibitem{Wood2025}
\bibinfo{author}{Wood, B.~M.} \emph{et~al.}
\newblock \bibinfo{title}{{UMA}: A family of universal models for atoms}
  (\bibinfo{year}{2025}).
\newblock \bibinfo{note}{{arXiv}:2506.23971}.

\bibitem{Solozhenko2001}
\bibinfo{author}{Solozhenko, V.~L.}, \bibinfo{author}{Andrault, D.},
  \bibinfo{author}{Fiquet, G.}, \bibinfo{author}{Mezouar, M.} \&
  \bibinfo{author}{Rubie, D.~C.}
\newblock \bibinfo{title}{Synthesis of superhard cubic {BC$_2$N}}.
\newblock \emph{\bibinfo{journal}{Appl. Phys. Lett.}}
  \textbf{\bibinfo{volume}{78}}, \bibinfo{pages}{1385--1387}
  (\bibinfo{year}{2001}).

\bibitem{Weisfeiler1968}
\bibinfo{author}{Weisfeiler, B.} \& \bibinfo{author}{Leman, A.}
\newblock \bibinfo{title}{A reduction of a graph to a canonical form and an
  algebra arising during this reduction}.
\newblock \emph{\bibinfo{journal}{Nauchno-Technicheskaya Informatsiya, Ser. 2}}
  \textbf{\bibinfo{volume}{9}}, \bibinfo{pages}{12--16} (\bibinfo{year}{1968}).

\bibitem{Shervashidze2011}
\bibinfo{author}{Shervashidze, N.}, \bibinfo{author}{Schweitzer, P.},
  \bibinfo{author}{van Leeuwen, E.~J.}, \bibinfo{author}{Mehlhorn, K.} \&
  \bibinfo{author}{Borgwardt, K.~M.}
\newblock \bibinfo{title}{Weisfeiler-{L}ehman graph kernels}.
\newblock \emph{\bibinfo{journal}{J. Mach. Learn. Res.}}
  \textbf{\bibinfo{volume}{12}}, \bibinfo{pages}{2539--2561}
  (\bibinfo{year}{2011}).

\bibitem{Kriege2020}
\bibinfo{author}{Kriege, N.~M.}, \bibinfo{author}{Johansson, F.~D.} \&
  \bibinfo{author}{Morris, C.}
\newblock \bibinfo{title}{A survey on graph kernels}.
\newblock \emph{\bibinfo{journal}{Appl. Netw. Sci.}}
  \textbf{\bibinfo{volume}{5}}, \bibinfo{pages}{6} (\bibinfo{year}{2020}).

\bibitem{XuGNN2019}
\bibinfo{author}{Xu, K.}, \bibinfo{author}{Hu, W.}, \bibinfo{author}{Leskovec,
  J.} \& \bibinfo{author}{Jegelka, S.}
\newblock \bibinfo{title}{How powerful are graph neural networks?}
\newblock \emph{\bibinfo{journal}{International Conference on Learning
  Representations}}  (\bibinfo{year}{2019}).
\newblock \bibinfo{note}{{arXiv}:1810.00826}.

\bibitem{Morris2019}
\bibinfo{author}{Morris, C.} \emph{et~al.}
\newblock \bibinfo{title}{Weisfeiler and {L}eman go neural: Higher-order graph
  neural networks}.
\newblock \emph{\bibinfo{journal}{Proc. AAAI Conf. Artif. Intell.}}
  \textbf{\bibinfo{volume}{33}}, \bibinfo{pages}{4602--4609}
  (\bibinfo{year}{2019}).

\bibitem{Xie2018}
\bibinfo{author}{Xie, T.} \& \bibinfo{author}{Grossman, J.~C.}
\newblock \bibinfo{title}{Crystal graph convolutional neural networks for an
  accurate and interpretable prediction of material properties}.
\newblock \emph{\bibinfo{journal}{Phys. Rev. Lett.}}
  \textbf{\bibinfo{volume}{120}}, \bibinfo{pages}{145301}
  (\bibinfo{year}{2018}).

\bibitem{Xu2022}
\bibinfo{author}{Xu, W.}, \bibinfo{author}{Reuter, K.} \&
  \bibinfo{author}{Andersen, M.}
\newblock \bibinfo{title}{Predicting binding motifs of complex adsorbates using
  machine learning with a physics-inspired graph representation}.
\newblock \emph{\bibinfo{journal}{Nat. Comput. Sci.}}
  \textbf{\bibinfo{volume}{2}}, \bibinfo{pages}{443--450}
  (\bibinfo{year}{2022}).

\bibitem{Griffiths2023}
\bibinfo{author}{Griffiths, R.-R.} \emph{et~al.}
\newblock \bibinfo{title}{{GAUCHE}: A library for {G}aussian processes in
  chemistry}.
\newblock \emph{\bibinfo{journal}{Advances in Neural Information Processing
  Systems}} \textbf{\bibinfo{volume}{36}}, \bibinfo{pages}{76923--76946}
  (\bibinfo{year}{2023}).

\bibitem{Seko2020}
\bibinfo{author}{Seko, A.} \& \bibinfo{author}{Ishiwata, S.}
\newblock \bibinfo{title}{Prediction of perovskite-related structures in
  {ACuO$_{3-x}$} ({A} = {Ca}, {Sr}, {Ba}, {Sc}, {Y}, {La}) using density
  functional theory and {B}ayesian optimization}.
\newblock \emph{\bibinfo{journal}{Phys. Rev. B}}
  \textbf{\bibinfo{volume}{101}}, \bibinfo{pages}{134101}
  (\bibinfo{year}{2020}).

\bibitem{Bartok2013}
\bibinfo{author}{Bart{\'o}k, A.~P.}, \bibinfo{author}{Kondor, R.} \&
  \bibinfo{author}{Cs{\'a}nyi, G.}
\newblock \bibinfo{title}{On representing chemical environments}.
\newblock \emph{\bibinfo{journal}{Phys. Rev. B}} \textbf{\bibinfo{volume}{87}},
  \bibinfo{pages}{184115} (\bibinfo{year}{2013}).

\bibitem{Drautz2019}
\bibinfo{author}{Drautz, R.}
\newblock \bibinfo{title}{Atomic cluster expansion for accurate and
  transferable interatomic potentials}.
\newblock \emph{\bibinfo{journal}{Phys. Rev. B}} \textbf{\bibinfo{volume}{99}},
  \bibinfo{pages}{014104} (\bibinfo{year}{2019}).

\bibitem{Prayogo2022}
\bibinfo{author}{Prayogo, G.~I.} \emph{et~al.}
\newblock \bibinfo{title}{{SHRY}: Application of canonical augmentation to the
  atomic substitution problem}.
\newblock \emph{\bibinfo{journal}{J. Chem. Inf. Model.}}
  \textbf{\bibinfo{volume}{62}}, \bibinfo{pages}{2909--2915}
  (\bibinfo{year}{2022}).

\end{thebibliography}
\end{document}